\documentclass[a4paper,11pt]{article}

\rightline{
	{\vbox {			
		\hbox{JLAB-THY-26-5038}
}}}
\vspace*{-0.6cm}

\usepackage{pos}

\title{Inclusive Heavy Quarkonium Production in Lepton-Hadron Scattering in a Joint QCD+QED Factorization Approach}
\ShortTitle{Inclusive quarkonium production in joint QCD+QED factorization}
\author[a,b]{Jian-Wei Qiu}
\author[c]{Kazuhiro Watanabe}
\author*[a]{Jia-Yue Zhang}
\affiliation[a]{Theory Center, Jefferson Lab, Newport News, Virginia 23606, USA}
\affiliation[b]{Department of Physics, William \& Mary, Williamsburg, Virginia 23187, USA}
\affiliation[c]{Department of Physics, Graduate School of Science, Tohoku University, Sendai 980-8578, Japan}
\emailAdd{jzhang@jlab.org}

\abstract{The Electron-Ion Collider (EIC) will provide an opportunity for precision studies of heavy quarkonium production in electron-proton ($ep$) and electron-nucleus ($eA$) collisions. Comparing production in $ep$ and $eA$ collisions provides a clean, well-controlled experimental environment to study interactions between a heavy quark pair and nuclear matter and the formation of quarkonia in a nuclear medium. In this talk, we present the first calculation of heavy quarkonium production in both $ep$ and $eA$ collisions at EIC energies using a joint QCD+QED factorization framework that treats collision-induced QCD and QED radiation on equal footing, and discuss the role of the nuclear medium in exploring how a physical quarkonium is emerged from a produced heavy quark pair. We also systematically extend our calculation to include next-to-leading power contributions in the inverse power expansion of quarkonium transverse momentum, and discuss the matching of our calculations to fixed-order calculations in the nonrelativistic Quantum Chromodynamics (NRQCD) approach.}

\FullConference{13th International Conference on Hard and Electromagnetic Probes of High-Energy Nuclear Collisions (HardProbes2026)\\
21-26 June 2026\\
Nashville, Tennessee, USA\\}

\begin{document}
\maketitle

\section{Introduction}

How a heavy quark pair (mass $m_Q\gg\Lambda_{\rm QCD}$) produced in a hard collision evolves into a physical quarkonium remains a central question in Quantum Chromodynamics (QCD). While heavy quark pair production is well understood within the NRQCD factorization approach~\cite{nrqcd}, a consistent description of quarkonium production and polarization remains elusive due to the lack of universality of the long-distance matrix elements (LDMEs), extracted from available data, which are needed to quantify how a quarkonium is emerged from a produced heavy quark pair according to NRQCD.

For quarkonium production at large transverse momentum ($p_T$), where $p_T \gg 2m_Q$, the cross section is primarily determined by the fragmentation of a single parton produced at leading power (LP) in the $1/p_T$ expansion, while short-distance heavy quark pair production followed by its fragmentation contributes at next-to-leading power (NLP)~\cite{kqs,kmqs}. Both contributions factorize into universal single- and double-parton fragmentation functions (FFs), which resum $\ln(p_T^2/m_Q^2)$-type collinear logarithms through evolution equations, with input distributions at $\mu_0 \sim 2m_Q$ calculable in NRQCD factorization~\cite{Ma:2013yla,Ma:2014eja}.

Early studies of hadronic collisions showed the importance of NLP contributions, including the logarithmically enhanced heavy quark pair contribution, at moderate $p_T$~\cite{Lee:2021oqr,lee}, but the input distributions at $\mu_0$ remain poorly constrained by the lack of precise data. Complementary to hadronic collisions, the EIC~\cite{eicwhite} will provide new opportunities to study quarkonium production in $ep$ and $eA$ collisions, with considerably cleaner initial states and different partonic channels. Since the beam electron primarily radiates photons, a consistent description of $ep$ and $eA$ collisions requires Quantum Electrodynamics (QED) alongside QCD. In this talk, we present single-inclusive quarkonium production at large $p_T$ in $ep$ and $eA$ collisions at EIC energies within a joint QCD+QED factorization framework~\cite{liu1,liu2}, retaining both LP single-parton and NLP double-parton fragmentation contributions.

\section{Joint QCD+QED factorization}

The joint QCD+QED factorization formalism for inclusive quarkonium ($H$) production in $ep$ collisions can be written as:
\begin{align}
 \frac{\mathrm{d}\sigma_{ep\to HX}}{\mathrm{d}^2p_T\,\mathrm{d}y}
 \simeq \sum_{i,j} f_{i/e}\otimes f_{j/p}\otimes
 \bigg[&\sum_{k} C_{ij\to k}\otimes D_{k\to H}+\sum_\kappa C_{ij\to[Q\bar Q(\kappa)]}\otimes {\cal D}_{[Q\bar Q(\kappa)]\to H}\bigg],
 \label{eq:joint}
\end{align}
where $C_{ij\to k}$ and $C_{ij\to[Q\bar Q(\kappa)]}$ are perturbatively calculable hard coefficients at LP and NLP, respectively. At the current precision, the index $i$ includes $e,\bar{e},\gamma$, the indices $j,k$ include $q,\bar{q},g$ with $q$ denoting light quarks, and $\kappa$ labels the heavy-quark pair's spin and color state. The $f_{i/e}$ and $f_{j/p}$ are universal lepton distribution functions (LDFs) and the parton distribution functions (PDFs), respectively. $D_{k\to H}$ and ${\cal D}_{[Q\bar Q(\kappa)]\to H}$ represent, respectively, LP and NLP quarkonium FFs, which are also universal. The LP and NLP quarkonium FFs satisfy a set of coupled QCD evolution equations~\cite{kmqsEvolution} that allow renormalization-group-improved factorization at NLP accuracy~\cite{kqs,kmqs}. In contrast, the LDFs obey the coupled QCD+QED evolution equations~\cite{Qiu:2026fed}. In our convention, positive rapidity is aligned with the electron beam's momentum direction.

Collision-induced QED radiation, as well as QCD radiation, were previously treated in Monte Carlo simulations. However, those studies typically imposed a cutoff on the exchanged photon's virtuality $Q^2$ to separate the direct-photon contribution ($Q^2\approx 0$) from the resolved-photon contribution, even though such an artificial cutoff can introduce uncertainties and require parameter tuning. Now the joint QCD+QED factorization approach provides a unified, systematic description of photoproduction and electroproduction without a cutoff, enabling us to handle collinear singularities arising along the directions of the beam electron, hadron, and observed quarkonium~\cite{liu1,liu2, Qiu:2026fed}.

\section{Single inclusive $J/\psi$ production at the EIC}

\begin{figure}[!t]
 \centering
 \includegraphics[width=0.9\textwidth]{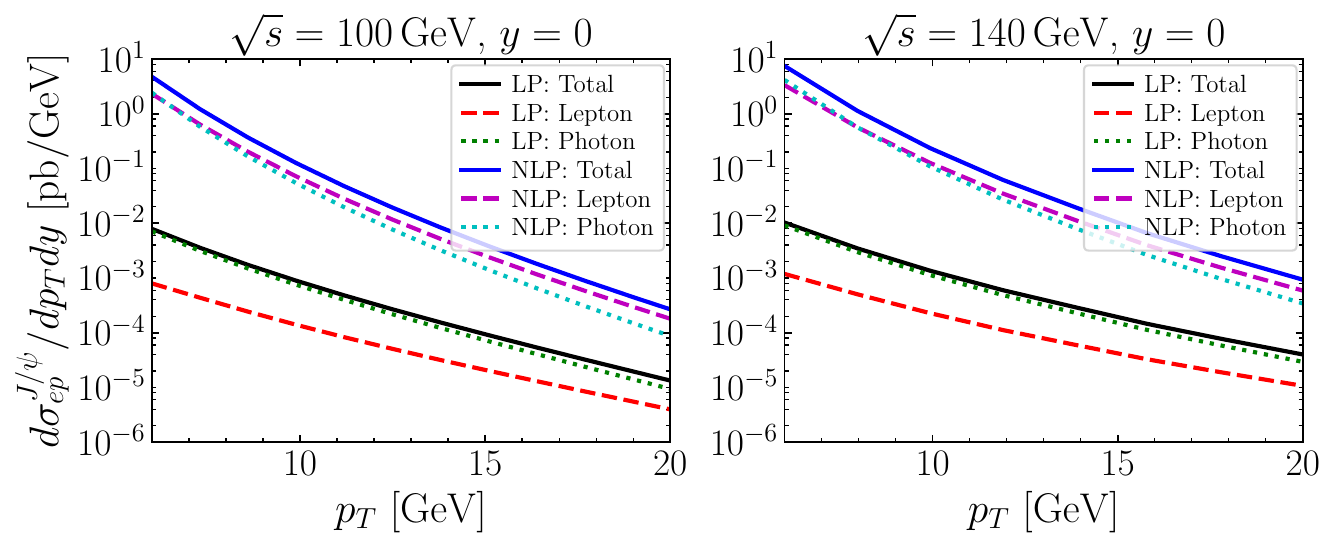}
 \caption{Inclusive $J/\psi$ production differential cross section at mid-rapidity in $ep$ collisions with $\sqrt{s}=100\,\mathrm{GeV}$ - current design energy (left) and $\sqrt{s}=140\,\mathrm{GeV}$ - future upgradable energy (right).}
 \label{fig:ep}
\end{figure}

We present in Fig.~\ref{fig:ep} numerical results for the $p_T$ spectrum of single-inclusive $J/\psi$ production in $ep$ collisions at EIC energies.
Within the joint QCD+QED factorization framework \eqref{eq:joint}, we employ the LDFs from Ref.~\cite{Qiu:2026fed} and the nCTEQ26 proton PDF set at NLO accuracy~\cite{nCTEQ:2026ewr}. The $J/\psi$ FFs are obtained by fitting LHC data on high-$p_T$ $J/\psi$ production~\cite{Lee:2021oqr,lee}. 

Unlike hadronic collisions at the LHC~\cite{Lee:2021oqr,lee}, the NLP contribution in Fig.~\ref{fig:ep} is much larger than the LP contribution across a wide range of $p_T$ due to several reasons: lower EIC energy, different partonic channels (fewer gluon jets to $J/\psi$), and steeply falling LP FFs as the momentum fraction $z$ increases, whereas NLP FFs peak at large $z$. Although the photoproduction channel dominates the LP term, both the leptoproduction and photoproduction channels contribute significantly at NLP, though their slopes differ slightly. Photoproduction is generally sizable at lower $p_T$. These results encourage us to investigate NLP contributions further.

\begin{figure}[!t]
 \centering
 \includegraphics[width=0.9\textwidth]{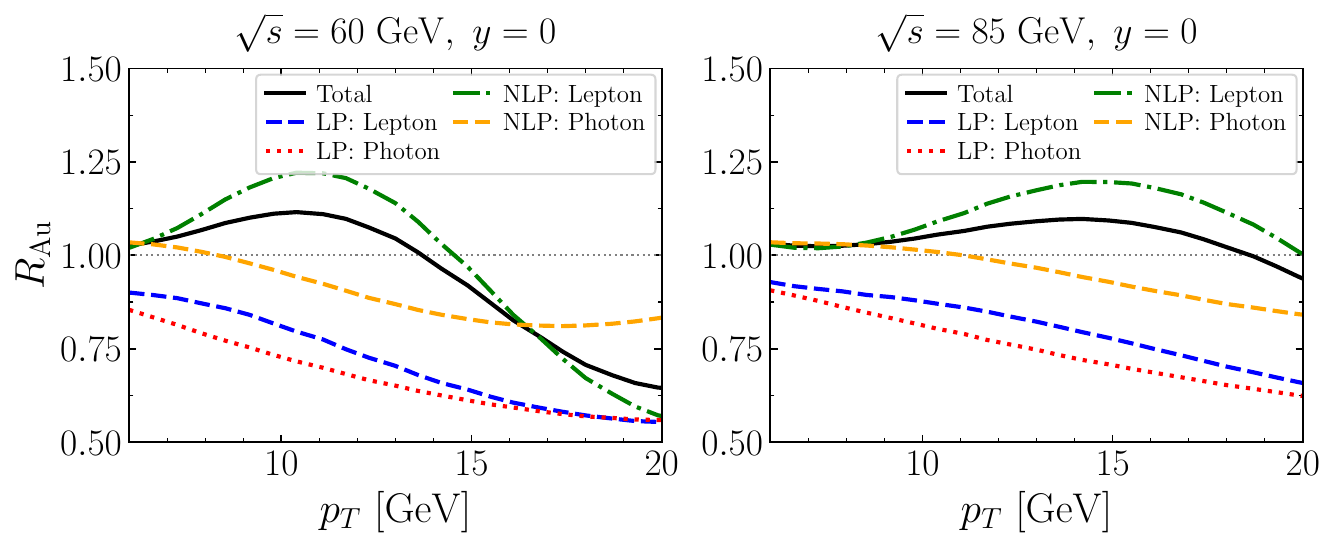}
 \caption{Ratios of inclusive $J/\psi$ production cross section in $e{\rm Au}$ collisions to that in $ep$ collisions scaled by atomic number. (Left): $\sqrt{s}=60\,\mathrm{GeV}$. (Right): $\sqrt{s}=85\,\mathrm{GeV}$.}
 \label{fig:nuclear}
\end{figure}

Next, we extend the framework to quarkonium production in $eA$ collisions by replacing proton PDFs with nuclear PDFs (nPDFs). At the same collision energy per nucleon, we study the nuclear modification factor defined as:
\begin{align}
 R_A(p_T,y)=\frac{1}{A}
 \frac{\mathrm{d}\sigma_{eA\to H X}/\mathrm{d}p_T\,\mathrm{d}y}
 {\mathrm{d}\sigma_{ep\to H X}/\mathrm{d}p_T\,\mathrm{d}y}.
 \label{eq:ra}
\end{align}
Figure~\ref{fig:nuclear} presents the nuclear modification factor for $J/\psi$ production in $e\mathrm{Au}$ collisions using the new nCTEQ26 nPDF set, and the total contribution (LP+NLP) is dominated by the NLP contribution, as shown in Fig.~\ref{fig:ep}. Since the NLP contribution is dominated by the partonic channel $e + g \to [c\bar{c}] + e$, the shape of $R_{\rm Au}$ in Fig.~\ref{fig:nuclear} exactly follows the modification of the gluon PDF in the nCTEQ26 set. On the other hand, light-quark-initiated partonic channels dominate the LP contributions, and the shape of $R_{\rm Au}$ from the LP contributions in Fig.~\ref{fig:nuclear} is consistent with the nuclear modification of light-quark distributions.

When the quarkonium's $p_T$ approaches the $2m_Q$ threshold, the $\ln(p_T^2/m_Q^2)$-type contributions no longer dominate the production cross-section. At the same time, $p_T\gtrsim 2m_Q \gg \Lambda_\mathrm{QCD}$ remains a hard scale for the colliding lepton and hadron. Therefore, we can calculate the heavy quark pair production using collinear factorization at fixed order (FO) and NRQCD factorization for the hadronization process, yielding the following matching condition:
\begin{align}
 \mathrm{d}\sigma^{\rm theory}  =\mathrm{d}\sigma^{\rm resum}_{\rm LP+NLP} +\mathrm{d}\sigma_{\rm NRQCD}^{\rm FO}  -\mathrm{d}\sigma^{\rm matching},
 \label{eq:matching}
\end{align}
where the first term is from \eqref{eq:joint} with resummation of large $\ln(p_T^2/m_Q^2)$-type corrections, the second term is the FO NRQCD contribution, and the third term subtracts the asymptotic overlap to avoid double counting. We will present a detailed study of this matching in future publications.

\section{Summary}

Heavy quarkonium production in $ep$ and $eA$ collisions, together with the joint QCD+QED factorization formalism and universal LDFs, PDFs, FFs, and LDMEs, should provide a well-controlled probe of the gluonic content of the nucleon and nucleus. Beyond nPDFs, nuclear modification factors could be further modified by multiple scattering of a heavy quark pair in the nuclear medium, providing rich information on the production mechanism of quarkonia~\cite{qvzSuppression}.

\section*{Acknowledgments}
The work of K.W. is supported by JSPS KAKENHI Grant No.~JP25K07286. The work of J.Q. and J.Z. is supported in part by the U.S. Department of Energy, Office of Science, Office of Nuclear Physics under Contract No.~89243126CSC000213.

\end{document}